\documentclass[conference]{IEEEtran}
\IEEEoverridecommandlockouts
\usepackage[utf8]{inputenc}
\usepackage[style=ieee,minnames=3,maxnames=3]{biblatex}
\usepackage{amsmath,amssymb,amsfonts}
\usepackage{algorithmic}
\usepackage{graphicx}
\usepackage{textcomp}
\usepackage{xcolor}
\def\BibTeX{{\rm B\kern-.05em{\sc i\kern-.025em b}\kern-.08em
    T\kern-.1667em\lower.7ex\hbox{E}\kern-.125emX}}

\usepackage[nolist,nohyperlinks]{acronym}
\usepackage{booktabs}
\usepackage[disable]{easy-todo}
\usepackage{mathtools}
\acrodef{mu}[MU]{machine units}
\acrodef{dag}[DAG]{directed acyclic graph}
\acrodef{artiq}[ARTIQ]{advanced real-time infrastructure for quantum physics}
\acrodef{nisq}[NISQ]{noisy intermediate-scale quantum}
\acrodef{dsl}[DSL]{domain-specific language}
\acrodef{jit}[JIT]{just-in-time}
\acrodef{rb}[RB]{randomized benchmarking}
\acrodef{dds}[DDS]{direct digital synthesis}
\acrodef{dac}[DAC]{digital-to-analog converter}
\acrodef{adc}[ADC]{analog-to-digital converter}
\acrodef{awg}[AWG]{arbitrary waveform generator}
\acrodef{fpga}[FPGA]{field-programmable gate array}
\acrodef{yb171}[${}^{171}$Yb$^+$]{Ytterbium 171}
\acrodef{pmt}[PMT]{photomultiplier tube}
\acrodef{api}[API]{application programming interface}
\acrodef{rpc}[RPC]{remote procedure call}
\acrodef{mw}[MW]{microwave}
\acrodef{bb1}[BB1]{broadband}
\acrodef{sk1}[SK1]{Solovay-Kitaev}
\acrodef{spam}[SPAM]{state preparation and measurement}
\acrodef{cw}[CW]{continuous wave}
\acrodef{rtio}[RTIO]{real-time I/O}
\acrodef{sqst}[SQST]{single-qubit state tomography}
\acrodef{gst}[GST]{gate set tomography}
\acrodef{ddb}[DDB]{device database}
\acrodef{rpc}[RPC]{remote procedure call}
\acrodef{ast}[AST]{abstract syntax tree}

\acrodef{dax}[DAX]{Duke ARTIQ extensions}
\acrodef{staq}[STAQ]{software-tailored architecture for quantum co-design}
\acrodef{rc}[RC]{red chamber}
\begin{document}

\title{Functional Simulation of\\Real-Time Quantum Control Software
}

\author{
\IEEEauthorblockN{Leon~Riesebos}
\IEEEauthorblockA{Department of Electrical and Computer Engineering\\
Duke University, NC 27708, USA\\
leon.riesebos@duke.edu}
\and
\IEEEauthorblockN{Kenneth~R.~Brown}
\IEEEauthorblockA{Department of Electrical and Computer Engineering\\
Duke University, NC 27708, USA\\
kenneth.r.brown@duke.edu}
}

\maketitle

\begin{abstract}
Modern quantum computers rely heavily on real-time control systems for operation.
Software for these systems is becoming increasingly more complex due to the demand for more features and more real-time devices to control.
Unfortunately, testing real-time control software is often a complex process, and existing simulation software is not usable or practical for software testing.
For this purpose, we implemented an interactive simulator that simulates signals at the \acl{api} level.
We show that our simulation infrastructure simulates kernels 6.9 times faster on average compared to execution on hardware, while the position of the timeline cursor is simulated with an average accuracy of 97.9\% when choosing the appropriate configuration.

% 100 words (not for QCE22?)

\end{abstract}

\begin{IEEEkeywords}
real-time control software,
signal simulation,
software testing,
quantum computing
\end{IEEEkeywords}

% Content
\acresetall
\section{Introduction}

State-of-the-art quantum hardware is becoming increasingly powerful with recent systems demonstrating computations on tens of qubits \cite{arute2019quantum, ryan2021realization, postler2021demonstration, wang201816, pogorelov2021compact, acharya2022suppressing, 810291}.
Recent papers \cite{arute2019quantum, kim2020hardware, blok2020quantum, pogorelov2021compact} have shown that such systems rely heavily on real-time control systems to control tens to hundreds of devices with nanosecond precision.
Programmable real-time control systems, as described in \cite{bourdeauducq_2016_51303, negnevitsky2018feedback, ioncontrol, fu2019eqasm, ryan2017hardware}, are already available and widely adopted. An often underexposed area of such real-time control systems is the increasingly complex control software required to operate them.
Larger quantum systems control more real-time devices, which leads to an increasing amount of software. In addition, real-time software is taking on more responsibilities ranging from hardware latency compensation to decomposing quantum gates into device control which further increases its complexity.

\begin{figure}
    \centering
    \includegraphics[scale=0.8]{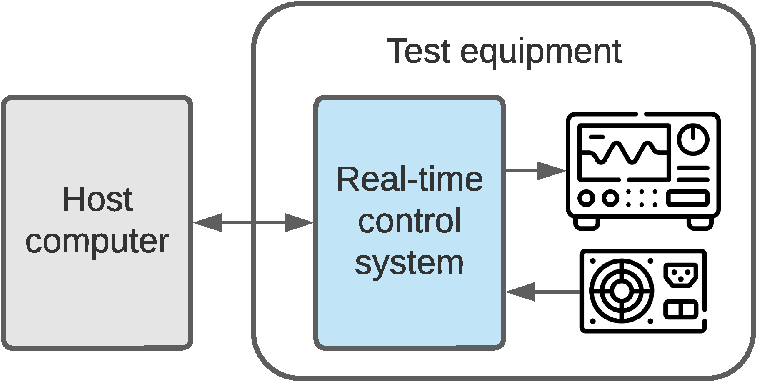}
    \caption{The equipment required for hardware testing, which includes the real-time control system, oscilloscopes, and signal generators.}
    \label{fig:test_equipment}
\end{figure}

With the growing complexity of real-time control software, functional testing and verification is becoming increasingly important.
Unfortunately, testing real-time control software is often complex, time-consuming, and resource-intensive. Testing on hardware requires access to control hardware and test equipment, such as oscilloscopes and signal generators, to probe and stimulate the control system, as illustrated in Figure~\ref{fig:test_equipment}. Even if all required test equipment is available, configuring the equipment to simulate the correct test signals can be complex and time-consuming. Additionally, black-box testing on hardware might not give enough insight into the state of the software if incorrect behavior is observed.
Software testing with hardware requires hardware to be available, which might not be the case in the early stages of development.
The use of simulation could enable testing of real-time control software, but simulators are usually not available for real-time control systems, as is the case for \cite{bourdeauducq_2016_51303, negnevitsky2018feedback, ioncontrol, fu2019eqasm}.
Existing simulation approaches that might be available, such as cycle-accurate hardware simulation, often focus on the microarchitectural level. Such simulations are too slow, inflexible, and low-level to be useful for testing real-time control software.

In this paper, we present an open-source functional simulator for real-time control software targeting the \ac{artiq} open-source software and hardware ecosystem \cite{bourdeauducq_2016_51303, Kasprowicz:20}.
Our interactive simulator simulates all aspects of real-time control software, including classical constructs, real-time events, and device input.
Real-time device signals are simulated at the \ac{api} level, which enables functional software testing and fast simulation speeds.
Our simulator integrates seamlessly into the \ac{artiq} host environment and is capable of simulating interactions between the host and the real-time control system.
With our simulation infrastructure, users can test and verify real-time control software using existing tools for step debugging, unit testing, and continuous integration.
Without the need for any of the test hardware shown in Figure~\ref{fig:test_equipment}, our simulator enables software testing in the early development stages.
We show that our kernel simulation is on average 6.9 times faster than execution on control hardware. Even with the presence of variable delays and simplified timing models for devices, the position of the timeline cursor is simulated with an average accuracy of 97.9\% when appropriately configured.

The remainder of this paper is structured as follows. Section~\ref{sec:related} briefly covers related work, and in Section~\ref{sec:overview} we will provide an overview of the \ac{artiq} hardware and software components that we will simulate. The design of our simulation platform is presented in Section~\ref{sec:simulation}, while the results of our performance and accuracy measurements can be found in Section~\ref{sec:evaluation}. We conclude our paper in Section~\ref{sec:conclusion}.

\section{Related work}
\label{sec:related}

% Keysight? I was hoping they had proprietary simulators for https://rfmw.em.keysight.com//wireless/helpfiles/m31xx_m33xxa_awg/Content/M3201A_M3202A_PXIe_AWG_Users_Guide/20%20Overview%20of%20Software%20and%20Programming%20Tools.html

% Labber simulator
% https://www.keysight.com/us/en/products/software/application-sw/labber-software.html

% https://auriga.com/blog/2020/simics-platform/

Real-time control hardware and software can be simulated with techniques similar to ones used for the simulation of embedded systems. 
Previous work such as \cite{rowson1994hardware, hines1997dynamic} proposes various techniques and approaches for such simulations.
Real-time control hardware can be simulated on a microarchitectural level based on their hardware description using the same binaries as the actual hardware. Cycle-accurate microarchitectural simulations can be performed with tools such as GEM5~\cite{gem5}, SystemC~\cite{systemc, systemc_reference}, Chisel~\cite{bachrach2012chisel}, or SimSoC~\cite{helmstetter2008simsoc}. Most of these tools can perform low-level and detailed cycle-accurate simulations of the hardware. Unfortunately, cycle-accurate simulations are often not usable for software testing and verification because simulations run slow and the simulated signals are too low-level for testing real-time software and device behavior.
These simulations also require detailed device models that might not be available in the early development stages.
The same holds for simulation techniques based on communication models of the microarchitecture, such as \cite{hines1997dynamic, erbas2007framework, pimentel2006systematic, helmstetter2008simsoc}.

High-level simulation approaches for quantum computer architecture as discussed in \cite{li2019sanq, fu2020quingo, riesebos2017pauli} can be fast and test real-time quantum programs. Unfortunately, these simulators operate on the quantum-gate level and do not simulate the real-time device control required to implement such operations.
Hence, high-level simulators are not usable for testing real-time control software on a real-time device and signal level.

\acresetall
\section{System overview}
\label{sec:overview}

\begin{figure}
    \centering
    \includegraphics[scale=0.8]{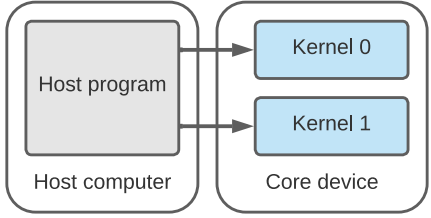}
    \caption{Schematic overview of the accelerator model with a host program and one or more kernels.}
    \label{fig:host_kernel}
\end{figure}

Our simulator targets the \ac{artiq} open-source software and hardware ecosystem \cite{bourdeauducq_2016_51303, Kasprowicz:20} which is used by dozens of research groups and has deployed over 200 real-time control systems worldwide.
The \ac{artiq} ecosystem combines a Python-based software environment with modular real-time control hardware, and its programming paradigm is based on the \emph{accelerator model} as described in \cite{riesebos2019quantum, fu2019eqasm, svore2018q, fu2020quingo, nguyen2020extending, smith2016practical, chong2017programming, stone2010opencl}.
The \ac{artiq} software environment runs on a host computer that communicates with the control hardware, also referred to as the \emph{core device}, over ethernet. Users can program the system using a Python host environment while kernels are executed on the core device as illustrated in Figure~\ref{fig:host_kernel}.

\subsection{Hardware}
\label{sub:hardware}

\begin{figure}
    \centering
    \includegraphics[scale=0.8]{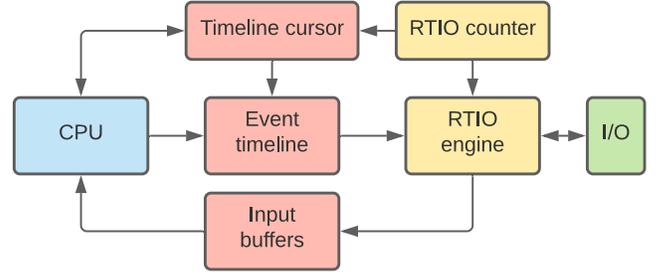}
    \caption{A schematic overview of the microarchitectural components in the core device.}
    \label{fig:hw_overview}
\end{figure}

The core device is driven by a \ac{fpga} which contains a classical CPU combined with an event-based \ac{rtio} subsystem similar to the systems outlined in \cite{fu2017experimental, fu2019eqasm}. Figure~\ref{fig:hw_overview} shows a simplified schematic of the relevant microarchitectural components in the \ac{fpga}.
The classical CPU will handle all classical instructions of the kernel and has additional access to a \emph{timeline cursor} and an \emph{event timeline}. The timeline cursor is a register that holds the current position on a timeline. The cursor is stored as an integer value that represents a time in \emph{\ac{mu}}, which normally corresponds to a timestamp expressed in nanoseconds. The CPU can also post events to the event timeline where an event is defined as a tuple of a timestamp and an I/O command. To change the state of a device, the CPU sets the timeline cursor to the time at which the change should occur before posting the I/O command to the event timeline. The current value of the timeline cursor will be used to store the event on the timeline. If the CPU posts two commands for the same device at the same timestamp, the last event will overwrite the first one.
By posting a series of events, a program can build up an event timeline that represents the real-time control of devices.

In parallel to the CPU's execution, the \ac{rtio} subsystem continuously verifies if any events are due. The \emph{\ac{rtio} counter} represents a timestamp in \ac{mu} and is incremented every nanosecond. The \emph{\ac{rtio} engine} reads the event timeline and verifies if any events are due based on the current value of the \ac{rtio} counter. If an event is due, the \ac{rtio} engine updates the corresponding device according to the command defined by the event.
In case an event generates a return value, for example, when reading the value of a digital input, the return value is inserted into the \emph{input buffers}. The CPU can read results from the input buffers whenever they are available.

For the \ac{rtio} system to operate properly, the \emph{slack} (i.e. the difference between the timeline cursor and the \ac{rtio} counter) must be positive. Posting an event with negative slack translates to changing the state of a device in the past, which is not possible. Doing so will result in an \emph{underflow exception}. Kernels normally start their program by synchronizing the timeline cursor to the \ac{rtio} counter and incrementing the timeline cursor with a fixed value of $125 \times 10^3$~\ac{mu} to ensure positive slack at the start of the program.

\subsection{Software}
\label{sub:software}

\begin{figure}
    \centering
    \includegraphics[scale=0.8]{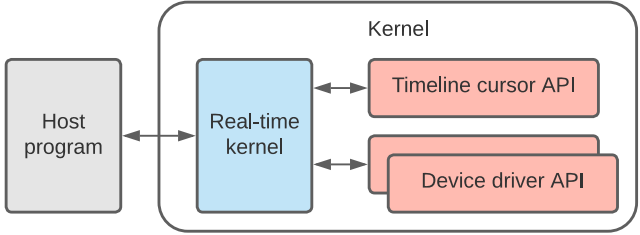}
    \caption{A schematic overview of a host program a kernel with access to \acp{api} for the timeline cursor and device drivers.}
    \label{fig:sw_overview}
\end{figure}

The \ac{artiq} software environment is Python-based and programs that run on the system are called \emph{experiments}. An experiment consists of Python code that runs on the host and can additionally contain kernel functions that run on the core device. Kernel functions are written in the \ac{artiq} \ac{dsl} which is a subset of the Python language. Inside kernels, programmers have access to additional functions to manipulate the timeline cursor, post events, and read input buffers. The latter two are normally not directly used by programmers as these functions are encapsulated in device drivers. Such device drivers provide an \ac{api} to translate functional device behavior (e.g. switch off a digital output pin) to low-level events.
A schematic overview of a host program and a kernel with access to \acp{api} for the timeline cursor and device drivers is shown in Figure~\ref{fig:sw_overview}.

When the host calls a kernel function, the \ac{artiq} compiler assembles a kernel binary at runtime which is then uploaded to and executed by the core device. Variables from the host environment that are accessed in a kernel will be compiled into the binary. During kernel execution, the host will handle any (a)synchronous \acp{rpc} initiated by the kernel. Once the kernel is finished executing, the context switches back to the host, and any variables modified in the kernel are synchronized with the host environment before the experiment resumes executing on the host. As a result, the context switch between host and kernel code is almost seamless from a programmer's perspective.

\section{Simulation}
\label{sec:simulation}

Our goal is to enable the simulation of real-time control software for software testing and verification. A simulator should integrate into the existing \ac{artiq} environment, simulate kernel execution, and simulate any interactions between the host environment and the kernel as described in Section~\ref{sec:overview}. The simulator should be fast enough to test complete experiments within a reasonable time. No real-time control hardware should be required to run simulations, only a model of the hardware listing the available devices.
Hardware/software co-simulation for embedded systems is not new, and existing work proposes various techniques and approaches for such simulations \cite{rowson1994hardware, hines1997dynamic}.
At the most detailed level, we find cycle-accurate simulations, such as \cite{gem5, systemc, bachrach2012chisel}, that take the same binary as the real system and simulate the components and registers of the microarchitecture in great detail. Such simulations require highly detailed models making them inflexible and potentially time-consuming to develop. Cycle-accurate simulators are extremely detailed and accurate but are also slow. It is not our goal to do performance analysis on the \ac{artiq} microarchitecture, and we do not need such a level of detail.
Since our target is software testing and not hardware performance analysis, we will focus on \ac{api} simulation. An \ac{api} simulation cross-compiles the target program to a simulator that implements the same \ac{api} as the target system. The simulator requires no execution model of the hardware and can therefore be fast.
Based on our requirements, we decide to target \emph{functional} simulation of kernels and real-time devices using \ac{api} simulation. Timeline cursor manipulations will be simulated at the \ac{api} level.
Real-time devices are simulated at their driver \ac{api} level, and functional behavior will be based on a simplified device model.
Hence, we will replace the timeline cursor \ac{api} and the device driver \acp{api} shown in Figure~\ref{fig:sw_overview} with calls to our simulation infrastructure.
The state of the \ac{rtio} counter and \ac{rtio} engine are not simulated, which would require the use of a cycle-accurate simulator.
Instead, we estimate the value of the \ac{rtio} counter when synchronizing the timeline cursor with the \ac{rtio} counter.

\begin{figure}
    \centering
    \includegraphics[scale=0.8]{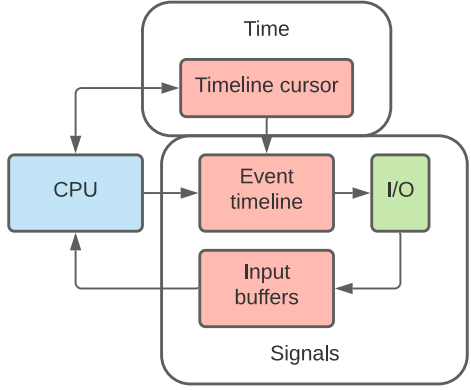}
    \caption{A schematic overview of the simulated microarchitectural components.}
    \label{fig:sim_overview}
\end{figure}

For simulation of real-time kernels, we will need to cover classical constructs (i.e. the CPU), the timeline cursor, the event timeline, and input buffers. Since both the host code and the classical constructs of the kernels are valid Python code, we decided to use the host Python process to simulate kernels. Hence, our simulator is implemented in Python and all components in Figure~\ref{fig:sw_overview} will be executed by the Python interpreter. Using the same Python process will also instantly implement host-kernel variable synchronization and handling of \acp{rpc}.
We decided to split the simulation of the remaining components into two parts: time and signals. The time component covers the simulation of the timeline cursor, and the signals component covers the simulation of the event timeline and input buffers. Figure~\ref{fig:sim_overview} shows a schematic overview of the simulated components. In the remainder of this section, we will cover time and signal simulation.

\subsection{Time}
\label{sub:time}

A kernel can read and write the value of the timeline cursor using the functions \verb;now_mu(); and \verb;at_mu(t);, respectively. Additionally, the cursor can be moved relative from its current position using the functions \verb;delay_mu(d); and \verb;delay(d);. The latter function is used with a delay time expressed in seconds instead of \ac{mu}. Since the delay in seconds is converted to a delay in \ac{mu}, the \verb;delay(d); function is not further discussed.
Functions used to modify the timeline cursor behave differently depending on the \emph{timing context} in which they are used. There are two timing contexts, \emph{sequential} and \emph{parallel}, which are used as regular Python context managers using the \verb;with; statement.
The two contexts are used to specify if a set of \ac{rtio} operations should be executed sequentially or in parallel. The contexts can be nested arbitrarily, and by default, every function starts in a sequential context. As a result, the timeline cursor simulation will have to adapt based on the current timing context.

In a sequential context, any modification to the timeline cursor is interpreted as a sequence of operations. Hence, two successive delays with duration $d_0$ and $d_1$ is equal to one delay with duration $d_0 + d_1$. Any call to \verb;at_mu(t); is applied instantly. Modifications to the timeline cursor in a parallel context are postponed such that operations in the context can be interpreted as parallel. When the program exits the parallel context, the timeline cursor will be moved forward by the duration of the longest positive delay.
If a parallel context containing delays with duration $d_0, \dots, d_n$ is entered with the timeline cursor at $t_{\textit{start}}$, the timeline cursor will be set to $t_{\textit{start}} + \max{(0, d_0, \dots, d_n)}$ when the context exits. In a parallel context, calls to \verb;at_mu(t); with value $t_{\textit{new}}$ are interpreted as delays with duration $t_{\textit{new}} - t_{\textit{start}}$.

We simulate the timeline cursor using a stack of simulation contexts that represent the nested timing contexts. The appropriate simulation context is pushed on and popped off the stack when a timing context is entered and exited, respectively. Each simulation context holds a current time $t_{\textit{current}}$ and a duration $t_{\textit{duration}}$ variable in \ac{mu}. When pushed to the stack, $t_{\textit{current}}$ is inherited from the simulation context currently at the top of the stack while $t_{\textit{duration}}$ is always initialized to zero. When a simulation context is popped off the stack, $t_{\textit{duration}}$ is propagated to the underlying simulation context as a delay.
There is a sequential and a parallel simulation context available and when the simulation starts, the stack is initialized with a sequential simulation context with $t_{\textit{current}} = 0$.
At any time, interactions with the timeline cursor are handled by the context at the top of the stack. \verb;now_mu(); always returns $t_{\textit{current}}$ while calls to \verb;delay_mu(d); are handled differently by the sequential and parallel simulation context. For a sequential simulation context, a delay with duration $d$ will increment $t_{\textit{current}}$ and $t_{\textit{duration}}$ by $d$ while for a parallel simulation context, $t_{\textit{current}}$ is not changed and $t_{\textit{duration}} = \max{(t_{\textit{duration}}, d)}$.
For both simulation contexts, calls to \verb;at_mu(t); with value $t_{\textit{new}}$ are converted to delays with duration $t_{\textit{new}} - t_{\textit{start}}$.
The described system using the stack of simulation contexts accurately simulates the behavior of the timeline cursor.

For correct synchronization of the timeline cursor to the \ac{rtio} counter, we keep track of a \emph{timeline horizon} which is essentially an estimation \ac{rtio} counter state.
For a simulation with events at timestamps $t_0, \dots, t_n$, the timeline horizon is defined as $\max{(t_\textit{cursor}, t_0, \dots, t_n)}$ where $t_\textit{cursor}$ is the current position of the timeline cursor.
When we synchronize the timeline cursor to the \ac{rtio} counter, we first set the position of the timeline cursor to the position of the timeline horizon before inserting a delay of $125 \times 10^3$~\ac{mu}.
Using the timeline horizon for synchronization is necessary to simulate code with negative delays correctly. Negative delays are commonly used to compensate for latencies of physical equipment.

\subsection{Signals}
\label{sub:signals}

For signal simulation, we need to simulate the event timeline and the input buffers. Interactions with the event timeline and input buffers happen through device drivers. We simulate device drivers on an \ac{api} level, and each driver simulates the signals and state of a device based on a simplified model. Signals will be simulated on a functional level, for example, frequency and phase for a \ac{dds} chip and a binary state for a digital output. To enable signal simulation, we will capture all function calls to drivers by replacing each device driver with a matching simulation driver.

During initialization, each simulation driver obtains one or more named signal objects corresponding to the state of the device. Each time a driver function is called to change the state of the device, the driver will \emph{push} new values to the appropriate signal objects. Pushing a new value to a signal object will cause an event to be created at the current position of the timeline cursor. Each signal object stores its events and therefore possesses a part of the complete event timeline of the system. If two events for a single signal have the same timestamp, the latest event overwrites the existing event.
Additionally, the simulation driver can keep an internal state and perform any additional processing for proper signal and time simulation.

To test real-time control software, we must have the ability to read the value of a signal at any given timestamp. To \emph{pull} the value of a signal at a specific timestamp, we search for the event with the highest timestamp that is less or equal to the timestamp of interest. The value of that event will represent the value of the signal at the given timestamp. If no event is found, the signal has not been set, and its value is unknown.

The last component that must be simulated is the input buffers. Values in these buffers originate from events with return values, such as sampling the value of a digital input device. For software testing, return values from input devices must be configurable by a test case. For that purpose, we introduce input signals that describe the state of a hypothetical device that generates the input signal observed by a device.
Just as output signals, input signals are obtained by the device drivers during initialization, for example, an input probability signal for a digital input device. When the simulation driver is called to sample the input value, the driver pulls the current value of the input probability signal and uses it to generate a return value. The return value is stored in the input buffer that is part of the simulation driver. Once the actual sampled value is requested from the driver, the value is taken from the buffer and returned.
Each input device has input signals that match the level of its functionality, such as input voltage for an \ac{adc} and input frequency for a digital edge counter.
During software testing, input signals can be configured using the same push/pull infrastructure used for output signals. This allows input signals to be adjusted using the same event timeline as output signals.

\subsection{Implementation}
\label{sub:implementation}

We have implemented a simulation platform for \ac{artiq} based on the proposed methodologies for time and signal simulation.
The simulator is part of our open-source library \ac{dax}
\cite{riesebos2021dax}  % Double blind
which integrates tightly with the \ac{artiq} open-source software environment.
The integration entry point for the \ac{dax} simulator is the \ac{ddb}, a central file in every \ac{artiq} project that defines the list of available real-time devices and their corresponding drivers. To enable simulation, users make a small modification that allows the \ac{dax} simulation infrastructure to mutate the \ac{ddb} before \ac{artiq} reads it at the start of an experiment. During \ac{ddb} mutation, all device drivers are replaced by matching simulation drivers, and an extra simulation configuration device is inserted into the \ac{ddb}. When the driver for the core device is loaded in an experiment, the core device simulation driver will be loaded, which in turn loads the driver for the simulation configuration device. The \ac{dax} simulation infrastructure is loaded during initialization of the simulation configuration device, which includes the setup of a time and a signal manager. Any other simulation drivers that are loaded will request their signal objects from the signal manager.

When the experiment runs and a kernel function is called, the core device driver is requested to compile the kernel and execute it on the core device. Instead, the simulation driver for the core device will just run the kernel function inside a sequential time context using the current Python process. Any interactions with the timeline cursor or time context \acp{api} are forwarded to the time manager for simulation while simulation drivers will perform all the signal simulations.
Events for each signal are stored in a sorted dictionary based on their timestamps, and binary search algorithms are used to push and pull events.

We integrated our simulation platform with the standard Python unit test framework such that users can run tests for real-time control software using existing testing environments. The \ac{dax} unit test base class, which inherits the standard Python unit test class, provides functions to push, pull, and test signal values at any timeline cursor position. 
Existing tools for step debugging, automated testing, and continuous integration will allow real-time control software to be tested to the same level as any other production-level software project.

\subsection{Limitations}
\label{sub:limitations}

Functional simulation of kernels at the \ac{api} level is fast and especially useful for testing and verification of real-time control software, but it also has limitations. Without simulation of the \ac{rtio} counter and the \ac{rtio} engine, slack can not be reliably simulated. As a result, \ac{api} simulation can not accurately predict underflow exceptions. A low-level and cycle-accurate microarchitectural simulation would be required to simulate slack. Such simulators are much slower and are not convenient for software testing and verification at the level discussed in this paper.

Some limitations are specific to our implementation of the simulation infrastructure. We use the running Python process to execute kernels, but the \ac{artiq} \ac{dsl} only supports a subset of the Python language. Hence, the simulation is more permissive than the \ac{artiq} compiler. We can mitigate this issue by compiling kernels before simulation. By default, the \ac{dax} simulator does not compile kernels to run simulations faster.

Host-kernel attribute synchronization also behaves differently in simulation. When running on a core device, the \ac{artiq} environment synchronizes host variables modified in a kernel when the kernel finished executing (see Section~\ref{sub:software}). During simulation, attributes are continuously synchronized due to the use of a single Python process for host and kernel code. The behavior of the simulator could be different when a kernel modifies the same variable used by an \ac{rpc} function it calls. Such code would have confusing semantics to start with, and we have not encountered any such code.

The model of the parallel timing context described in Section~\ref{sub:time} differs slightly from the timing model implemented in the \ac{artiq} compiler. The \ac{dax} simulator propagates the parallel semantics until a sequential context is entered (deep parallel) while the \ac{artiq} compiler only propagates the parallel semantics to top-level statements in the context (shallow parallel). Kernel code that potentially behaves differently with deep and shallow parallel semantics can be detected using \ac{ast} analysis. We have developed a separate tool
\cite{riesebos2020flake8}  % Double blind
that flags such kernel code.

\section{Evaluation}
\label{sec:evaluation}

\begin{table}
\centering
\begin{tabular}{@{}ll@{}}
\toprule
Label       & Experiment                          \\
\midrule
mw\_freq    & Microwave frequency scan            \\
mw\_rabi    & Microwave Rabi frequency scan       \\
mw\_ramsey  & Microwave Ramsey scan               \\
mw\_gate    & Microwave repeated gate scan        \\
gco\_freq   & Global co-propagating frequency scan \\
gco\_rabi   & Global co-propagating Rabi frequency scan \\
gco\_ramsey & Global co-propagating Ramsey scan   \\
ico\_freq   & Individual co-propagating frequency scan \\
ico\_ttime  & Individual co-propagating time scan \\
state\_init & Qubit state initialization scan     \\
tickle      & Tickle scan                         \\
direct\_rb  & Direct \acl{rb}                     \\
gst         & \Acl{gst}                           \\
sqst        & \Acl{sqst}                          \\
\bottomrule
\vspace{1pt}
\end{tabular}
\caption{List of experiments used for the evaluation.}
\label{tab:experiments}
\end{table}

To evaluate the performance of the \ac{dax} simulation platform, we measured its kernel execution time and compared it to the execution time on hardware.
We used two experimental trapped-ion quantum processors for our evaluation, the \ac{staq} system
\cite{kim2020hardware}  % Double blind
and the \ac{rc} system
\cite{wang2020high}  % Double blind
.
Both systems are controlled by an \ac{artiq} control system, but \ac{staq} uses a core device based on the Kasli~2.0 controller \cite{Kasprowicz:20} while \ac{rc} uses a KC705-based controller \cite{kc705}.
Besides the different real-time control systems and devices, the main difference between these two setups is that \ac{staq} is at cryogenic temperatures while \ac{rc} is at room temperature.
We chose 14 commonly used experiments with a single kernel for the \ac{staq} system.
The set of experiments, listed in Table~\ref{tab:experiments}, contains 11 scanning-type experiments used for calibration and three benchmarking experiments including, Direct \ac{rb} \cite{magesan2011scalable, PhysRevLett.123.030503, epstein2014investigating}, \ac{gst} \cite{blume2013robust}, and \ac{sqst} \cite{schmied2016quantum}.
Both systems use modular real-time control software developed with the \ac{dax} modular software framework \cite{riesebos2022modular}, and parts of the system-specific control software are available in the \ac{dax}-zoo repository~\cite{riesebos2022daxzoo}.
The three benchmark experiments are portable and can also run on \ac{rc} while the four \ac{mw} calibration experiments have an equivalent implementation for the \ac{rc} system.
All scanning-type experiments scan over 20 points and take 100 samples per point. Direct \ac{rb} is performed with circuit lengths starting at 1 and scaling up exponentially to 16. For each circuit length, we benchmark ten different circuits with 100 samples for each circuit. The \ac{gst} benchmarks are performed with a total of 523 different circuits based on our germs, taking 100 samples per circuit. Finally, \ac{sqst} is performed with a grid of 5 times 10 angles taking 100 samples for each point.

\begin{figure}
    \centering
    \includegraphics[width=\linewidth]{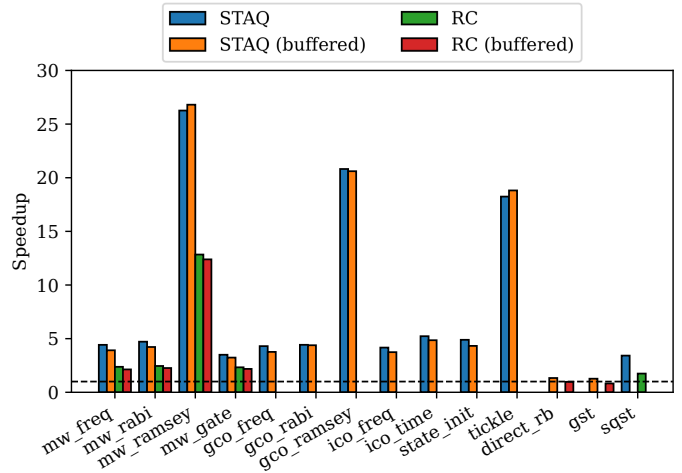}
    \caption{Kernel execution time speedup for our simulator relative to the execution time on a core device.}
    \label{fig:sim_speedup}
\end{figure}

For our evaluation, we run the experiments for both systems on a Kasli~2.0 controller.
The \ac{rc} software can run on an appropriately configured Kasli controller by replacing the \ac{ddb}.
All calibration experiments are executed with and without buffering. Buffering allows the real-time control software to schedule the operations for the next samples while the incoming data of earlier samples are kept temporally in hardware buffers. \Ac{artiq} supports such hardware buffers, but the real-time software must be designed appropriately to utilize them. Buffering can further increase the throughput and performance of kernels by reducing stalling time at the cost of increased latency between receiving and processing input events. None of the experiments are sensitive to the increased latency and will benefit from increased throughput. We configure a buffer size of 16 samples, which should be large enough to get the maximum performance gain achievable with buffering.
The Direct \ac{rb} and \ac{gst} experiments are always buffered with a fixed buffer size of 1 and \ac{sqst} is always unbuffered.
The kernel execution time is measured with nanosecond precision using the real-time clock available in the Kasli controller.
We then run the same experiments using our \ac{dax} simulation platform on a computer equipped with an AMD~Ryzen~7~3700X CPU and 32~GB of memory. The computer runs on Ubuntu~20.04~LTS, and the execution time of the kernel simulation is measured in nanoseconds using the standard Python time library.
All experiments run five times on hardware and five times in simulation to take the average simulation time.
Our measurements are performed using \ac{artiq} version~6.7659.c6a7b8a8 and the results are presented in Figure~\ref{fig:sim_speedup}.

% speedup PQA [ 4.4  4.7 26.3  3.5  4.3  4.4 20.8  4.2  5.2  4.9 18.2  3.4] 8.7
% speedup PQA (buffered) [ 3.9  4.2 26.8  3.2  3.8  4.4 20.6  3.7  4.9  4.3 18.8  1.3  1.3] 7.8
% speedup GT [ 2.4  2.5 12.8  2.3  1.7] 4.3
% speedup GT (buffered) [ 2.1  2.3 12.4  2.2  1.   0.8] 3.5
% speedup all 6.9

The results in Figure~\ref{fig:sim_speedup} show that simulation speeds up execution up to 26.8 times with an average speedup of 6.9 times.
Especially the mw\_ramsey, gco\_ramsey, and tickle experiments achieve large speedups. The exceptional speedup for these experiments is caused by the long delays that are part of the experiment. The core device waits for these delays before the kernel finishes execution, while the simulator only simulates the passing of time but does not wait for it.
The experiments that show the least speedup are the direct\_rb and gst experiments. For \ac{staq}, both experiments only yield a 1.3 times speedup, while for \ac{rc}, the direct\_rb experiment has no speedup and the gst experiment is slower with a speedup of 0.8 times.
The limited speedup of these two experiments is caused by short delays and a high number of operations, which results in a high event density.
As a result, the simulator must process many events while the experiment has a relatively short execution time on hardware.
In general, we could state that the execution time on hardware $t_\textit{hardware}$ is mostly limited by the length of delays inserted during the experiment. These delays sum up to the total length of the timeline and therefore the duration of the experiment when running on hardware. The execution time of the simulator $t_\textit{sim}$ is not much affected by delays and instead is mostly limited by the total number of events present in the experiment. We know that speedup is defined as $S = t_\textit{hardware} / t_\textit{sim}$. Roughly speaking, we can derive that the total duration of an experiment is proportional to speedup while the total number of events is inversely proportional to speedup.

We can see from Figure~\ref{fig:sim_speedup} that the experiments running on the \ac{rc} system always yield lower speedup compared to the same experiment running on \ac{staq}. The different results are caused by differences in the control for the cooling and pumping procedures. Both procedures are executed by all experiments at the start of each sample.
\ac{staq} uses three digital outputs and one \ac{dds} while \ac{rc} has additional features and uses five digital outputs and a \ac{dds}. As a result, \ac{rc} inserts more events for each cooling and pumping procedure. Additionally, \ac{staq} uses a constant \ac{dds} frequency for both procedures while \ac{rc} uses a different frequency for each procedure which adds two additional \ac{dds} configuration events for each sample. Hence, the total number of events for \ac{rc} experiments is higher than for \ac{staq} which reduces the speedup. The additional \ac{dds} operations also insert extra delays into the experiment, but these delays do not compensate for the increased number of events.
Figure~\ref{fig:sim_speedup} also shows buffered experiments tend to have slightly less speedup compared to their unbuffered counterparts. Buffering can reduce the execution time overhead of experiments resulting in faster execution on hardware. The total number of events per experiment is not affected by buffering. The result is a reduced speedup for experiments with buffering. The reduction in execution time by buffering is limited though due to the highly optimized control software.

In addition to speedup, we have also measured the timing accuracy of the simulated timeline cursor compared to execution on the core device.
High timing accuracy is not a specific requirement for correct functional simulation, but a simulator with high timing accuracy could be used for estimating the timing of experiments.
The timeline cursor simulation is accurate, but variable delays and inaccurate delays in simulated device drivers can still introduce errors.
Variable delays mainly occur when the timeline cursor is synchronized with the \ac{rtio} counter. Such synchronization is performed at least once at the start of the experiment (see Section~\ref{sub:hardware}) but can also occur at other moments. We simulate the synchronization of the timeline cursor using a timeline horizon and insert an additional delay of $125 \times 10^3$~\ac{mu}.
We would like to emphasize that the presence of a variable delay indicates that the relative timing between the events before and after the delay is not relevant, and any variation will not negatively impact the functionality of the experiment or the simulation. Hence, simulating timeline cursor synchronization with a timeline horizon is sufficient for correct functional simulation.
A variable delay can also occur when an experiment needs to wait for an input event that occurs at an unpredictable time, though none of the experiments in Table~\ref{tab:experiments} contain such constructions.
Inaccurate delays in simulated device drivers are often caused by a simplified timing model of the device driver. In practically all cases with inaccuracy, the simulated driver inserts less delay than the actual driver.

\begin{figure}
    \centering
    \includegraphics[width=\linewidth]{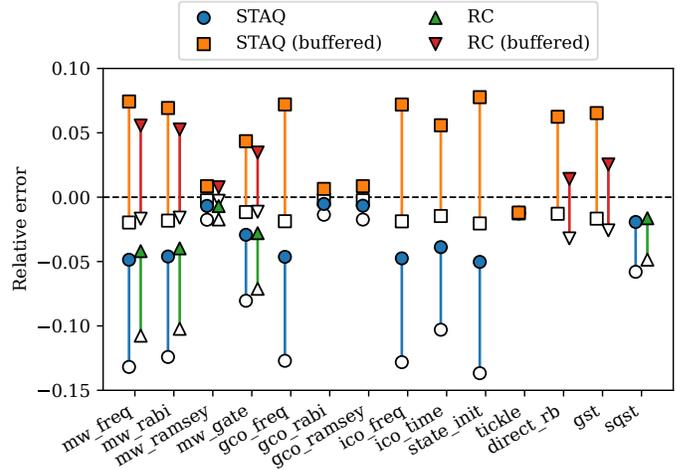}
    \caption{The error of the simulated timeline cursor relative to the timeline cursor of the core device. The filled markers represent the regular configuration while empty markers represent the optimistic configuration.}
    \label{fig:sim_error}
\end{figure}

To measure the timing accuracy of the simulated timeline cursor, we store the value of the timeline cursor after the first synchronization with the \ac{rtio} counter and at the end of the experiment. The difference between the two values represents the total length of the event timeline in \ac{mu}. We run the simulations with two configurations: regular and optimistic. When the timeline cursor is synchronized with the \ac{rtio} counter, our simulator inserts a fixed delay of $125 \times 10^3$ and $0$~\ac{mu} for the regular and optimistic configuration, respectively. We measured the event timeline length on the core device and with the two simulation configurations for all experiments listed in Table~\ref{tab:experiments} using the \ac{staq} and \ac{rc} system.
For each combination of system, experiment, and configuration, we calculate the relative error of the simulation which is defined as $(t_\textit{sim} - t_\textit{exe}) / t_\textit{exe}$ where $t_\textit{exe}$ and $t_\textit{sim}$ are the measured event timeline lengths on the core device and during simulation, respectively.
The results for are shown in Figure~\ref{fig:sim_error} and are also listed in Table~\ref{tab:sim_error_staq} and~\ref{tab:sim_error_rc}.

\begin{table}
\centering
\begin{tabular}{@{}lrrrr@{}}
\toprule
Experiment  & \multicolumn{2}{c}{\acs{staq}} & \multicolumn{2}{c}{\acs{staq} (buffered)} \\
            & Regular   & Optimistic   & Regular         & Optimistic        \\
\midrule
mw\_freq    & -4.9\%     & -13.2\%       & 7.4\%            & -2.0\%             \\
mw\_rabi    & -4.6\%     & -12.4\%       & 6.9\%            & -1.8\%             \\
mw\_ramsey  & -0.7\%     & -1.8\%        & 0.9\%            & -0.2\%             \\
mw\_gate    & -2.9\%     & -8.0\%        & 4.4\%            & -1.2\%             \\
gco\_freq   & -4.6\%     & -12.7\%       & 7.2\%            & -1.9\%             \\
gco\_rabi   & -0.5\%     & -1.4\%        & 0.6\%            & -0.2\%             \\
gco\_ramsey & -0.7\%     & -1.7\%        & 0.9\%            & -0.2\%             \\
ico\_freq   & -4.8\%     & -12.8\%       & 7.2\%            & -1.9\%             \\
ico\_time   & -3.9\%     & -10.3\%       & 5.6\%            & -1.5\%             \\
state\_init & -5.0\%     & -13.7\%       & 7.8\%            & -2.0\%             \\
tickle      & -1.2\%     & -1.3\%        & -1.2\%           & -1.3\%             \\
direct\_rb  &           &              & 6.2\%            & -1.3\%             \\
gst         &           &              & 6.5\%            & -1.7\%             \\
sqst        & -1.9\%     & -5.8\%        &                 &                  \\
\bottomrule
\vspace{1pt}
\end{tabular}
\caption{The error of the simulated timeline cursor relative to the timeline cursor of the core device for \acs{staq}.}
\label{tab:sim_error_staq}
\end{table}

\begin{table}
\centering
\begin{tabular}{@{}lrrrr@{}}
\toprule
Experiment  & \multicolumn{2}{c}{\acs{rc}} & \multicolumn{2}{c}{\acs{rc} (buffered)} \\
            & Regular  & Optimistic  & Regular        & Optimistic       \\
\midrule
mw\_freq    & -4.2\%    & -10.8\%      & 5.6\%           & -1.7\%            \\
mw\_rabi    & -4.0\%    & -10.2\%      & 5.3\%           & -1.6\%            \\
mw\_ramsey  & -0.7\%    & -1.7\%       & 0.8\%           & -0.3\%            \\
mw\_gate    & -2.8\%    & -7.1\%       & 3.5\%           & -1.1\%            \\
% gco\_freq   &          &             &                &                  \\
% gco\_rabi   &          &             &                &                  \\
% gco\_ramsey &          &             &                &                  \\
% ico\_freq   &          &             &                &                  \\
% ico\_time   &          &             &                &                  \\
% state\_init &          &             &                &                  \\
% tickle      &          &             &                &                  \\
direct\_rb  &          &             & 1.4\%           & -3.2\%            \\
gst         &          &             & 2.5\%           & -2.6\%            \\
sqst        & -1.6\%    & -4.9\%       &                &                 \\
\bottomrule
\vspace{1pt}
\end{tabular}
\caption{The error of the simulated timeline cursor relative to the timeline cursor of the core device for \acs{rc}.}
\label{tab:sim_error_rc}
\end{table}

% sim_error_2_scatter PQA [-0.049 -0.046 -0.007 -0.029 -0.046 -0.005 -0.007 -0.048 -0.039 -0.05 -0.012 -0.019] 0.03
% sim_error_2_scatter PQA_lb [-0.132 -0.124 -0.018 -0.08  -0.127 -0.014 -0.017 -0.128 -0.103 -0.137 -0.013 -0.058] 0.079
% sim_error_2_scatter PQA (buffered) [ 0.074  0.069  0.009  0.044  0.072  0.006  0.009  0.072  0.056  0.078 -0.012  0.062  0.065] 0.048
% sim_error_2_scatter PQA (buffered)_lb [-0.02  -0.018 -0.002 -0.012 -0.019 -0.002 -0.002 -0.019 -0.015 -0.02 -0.013 -0.013 -0.017] 0.013
% sim_error_2_scatter GT [-0.042 -0.04  -0.007 -0.028 -0.016] 0.027
% sim_error_2_scatter GT_lb [-0.108 -0.102 -0.017 -0.071 -0.049] 0.069
% sim_error_2_scatter GT (buffered) [0.056 0.053 0.008 0.035 0.014 0.025] 0.032
% sim_error_2_scatter GT (buffered)_lb [-0.017 -0.016 -0.003 -0.011 -0.032 -0.026] 0.017
% sim_error_2_scatter all_regular 0.036
% sim_error_2_scatter all_lb 0.044
% sim_error_2_scatter all_nbuf_regular 0.029
% sim_error_2_scatter all_buf_lb 0.014
% sim_error_2_scatter all_realistic 0.021
% t difference labels ['mw_freq', 'mw_rabi', 'mw_ramsey', 'mw_gate', 'sqst']
% t difference regular [0.268 0.25  0.042 0.186 0.204] 0.19
% t difference labels ['mw_freq', 'mw_rabi', 'mw_ramsey', 'mw_gate', 'direct_rb', 'gst']
% t difference buffered [0.3   0.278 0.042 0.197 0.721 0.601] 0.357
% t difference mean 0.281

The results in Figure~\ref{fig:sim_error} show the error of the simulated timeline cursor relative to the timeline cursor of the core device. The regular and optimistic configurations are represented by the filled and empty markers, respectively.
When comparing the results of the two different configurations, we see that the optimistic configuration always estimates a shorter timeline length, which is expected.
If we only look at the results for the optimistic configuration, we see that all have a relative error lower or equal to 0.0. The optimistic configuration represents the lower-bound execution time where variable delays are always zero. When running on actual hardware, variable delays are not always zero, and as a result, the optimistic configuration underestimates the timeline length.
We also noticed that all unbuffered results with regular configuration have a relative error lower or equal to 0.0. When running on hardware without buffers, the system has negative slack after each sample, and timeline synchronizations will insert delays larger than $125 \times 10^3$~\ac{mu}. The regular configuration underestimates the length of the variable delay and therefore underestimates the total timeline length. Regardless, the estimation of the regular configuration is better than that of the optimistic configuration for unbuffered experiments.
The opposite is true for buffered experiments. Buffering reduces the length of variable delays caused by timeline synchronizations by maintaining slack between samples. The regular configuration is often too pessimistic for buffered experiments and the estimation of the optimistic configuration is better most of the time.

We noticed two other trends in Figure~\ref{fig:sim_error} that relate to the total timeline length of experiments.
First, the results of some experiments have little spread, in particular mw\_ramsey, gco\_rabi, gco\_ramsey, and tickle. These are all calibration experiments with relatively long delays and long total timeline lengths. The long timeline length combined with the limited sources of errors (i.e. low density of variable delays and events) results in a small relative error and therefore, a small spread between different configurations.
Second, the results of the \ac{rc} system tend to be closer to 0.0 than the equivalent \ac{staq} results. We already mentioned that due to differences in the cooling and pumping procedures, the \ac{rc} system inserts more events for each sample of the experiment. These additional events also insert extra delays into the experiment. As a result, the total timeline length of \ac{rc} experiments are on average 28.1\% longer compared to their \ac{staq} equivalents. Again, the increased timeline length with no additional sources of errors reduces the relative error.

% To compute the average error, we take the absolute value of each relative error before calculating the average error.
Overall, the average relative error for the regular configuration is 3.6\%, and for the optimistic configuration, the average relative error is 4.4\%.
Based on our analysis of the regular and optimistic configurations, we concluded that the timeline length of buffered and unbuffered experiments are better estimated by the regular and optimistic configurations, respectively.
When choosing the optimistic configuration for buffered experiments and the regular configuration for unbuffered experiments, the resulting average relative error is reduced to 2.1\%, leading to an average accuracy of 97.9\%.
We can conclude that even in the presence of variable delays and simulated device drivers with simplified timing models, the position of the timeline cursor is simulated with high accuracy when choosing the appropriate configuration.

\acresetall
\section{Conclusion}
\label{sec:conclusion}

We have presented a functional simulation platform for real-time control software that enables software testing and verification.
To simplify testing and verification, timeline manipulations and device drivers are simulated on the \ac{api} level.
Our simulation platform accurately simulates a timeline cursor using a stack while the event timeline is simulated using signals and events.
Input signals are also simulated on a functional level and use the same interactive signal and event infrastructure used for output signals.
We implemented a simulator based on the proposed concepts, which is part of our open-source library \ac{dax}.
Our simulator integrates tightly into the \ac{artiq} environment and is capable of simulating real-time kernels and host-kernel interactions.
We integrated our simulator with the standard Python unit test frameworks such that real-time control software can be tested using existing tools for step debugging, unit testing, and continuous integration.
Compared to kernel execution on the core device, kernel simulation is 6.9 times faster on average.
Even with the presence of variable delays and simplified timing models for device drivers, the position of the timeline cursor is simulated with an average accuracy of 97.9\% when choosing the appropriate configuration.

\section*{Acknowledgment}
This work is funded by EPiQC, an NSF Expeditions in Computing (1832377), the Office of the Director of National Intelligence - Intelligence Advanced Research Projects Activity through an ArmyResearch Office contract (W911NF-16-1-0082) and the NSF STAQ project (1818914).

\printbibliography

\end{document}